\documentclass[12pt]{article}
\usepackage{amsmath}
\begin{document}
\vspace*{3cm}
\begin{center}

\textbf{\LARGE The Green function for thermal waves induced by attosecond laser 
pulses in one and three dimensions}

\vspace{2cm}

{\large Janina Marciak-Koz{\l}owska\\
Miros{\l}aw Koz{\l}owski}\footnote{Corresponding author, e-mail: MiroslawKozlowski@aster.pl}

\vspace{2cm}

Institute of Electron Technology, Al. Lotnik\'{o}w 32/46, 02-668 
Warsaw, Poland

\end{center}

\newpage
\begin{abstract}
In this paper the solution of the hyperbolic Klein-Gordon thermal 
equation are obtained and discussed. The analytical form of the 
solution -- Green functions are calculated for one and three dimensional 
cases. It is shown that only in three dimensional case the undisturbed, 
with one value of the velocity, thermal wave can be generated 
by attosecond laser pulses. The conductivity for the space-time 
inside the atom is calculated and the value 
$\sigma _{0} =10^{6} \frac{1}{\Omega {\rm m}} $
 is obtained.

\end{abstract}
\newpage

\section{Introduction}
The continuing progress in the technology of ultra-short pulse 
lasers has led to the landmark breakthrough and to the birth 
of attosecond physics~\cite{1}-\cite{6}. Attosecond pulses of X-rays are 
produced when intense sub-fs laser pulses interact with gases. 
Under the action of strong laser fields on the atom, a wave packet 
is formed each time the laser field passes its maximum value. 
Within the next laser period there is a probability that the 
electron having a high kinetic energy return to the ion. The 
energy of electrons which do return back to the ion is easily 
calculated from the classical electron trajectory in the laser 
field. For the maximum energy one obtains 
$E_{\max } \approx 3.2U_{P} $, where 
$U_{P} $ is the laser ponderomotive potential. Reencountering the ion, 
the electron radiates a burst of X-rays with energy up to 
$\hbar \omega _{c} =3.2U_{P} +I_{P} $, where 
$I_{P} $
 is the ionization potential. From the mechanism of electron-ion 
reencounter, one expects that the high harmonics are phase locked 
and merge in the form of the attosecond pulses. If the lasers 
pulse is many oscillations long, a train of attosecond pulses 
is produced. When the laser pulse duration is comparable to its 
period then solitary attosecond pulses can be generated.

In this paper the interaction of attosecond laser pulses with 
matter will be investigated. Considering the results presented 
is our monograph{\nobreakspace}[7] we will use the hyperbolic Klein-Gordon 
thermal equation (KGT){\nobreakspace}[7]. The solution of KGT in the 
form of the Green functions will be obtained and investigated.
\section{The model}
In the last few decades there has been extensive theoretical, 
and more recently experimental research on the quantum-classical 
transition. It has been noted that every physical system is, 
in fact an open quantum system interacting with its environment. 
Consequently the evolution of the system leads to the suppression 
of the quantum coherent effects. This process of environment 
induced decoherence has been considered to be an essential ingredient 
of the quantum classical transition.

In the book~\cite{7} the model for quantization of classical 
thermal field was proposed. As the result the quantum heat transport 
equation was obtained~\cite{7}  

    \begin{equation}
    \frac{1}{v^{2} } \frac{\partial ^{2} T}{\partial t^{2} } +\frac{m}{\hbar} \frac{\partial T}{\partial t} +\frac{2Vm}{\hbar ^{2} } T=\nabla ^{2} T.\label{eq1}
    \end{equation}
In equation~(\ref{eq1}) \textit{v} is the speed of the interaction propagation, 
$v=\alpha c$  where 
$\alpha $ is the fine structure constant, \textit{m} is the heaton mass, \textit{V} 
is potential and \textit{T} denotes temperature of the system, 
$\hbar $  is the Planck constant. In the procedure of the quantization 
the formula for discrete value of the quantum of the relaxation 
time was obtained:
    \begin{equation}
    \tau =\frac{\hbar }{mv^{2} } =\frac{\hbar }{m\alpha ^{2} c^{2} } .\label{eq2}
    \end{equation}
With the Eq.~(\ref{eq1}) and formula~(\ref{eq2}) the interatomic 
transport phenomena can be described. For the hydrogen type of 
atoms (i.e. with one electron active) 
$\tau $  is of the order of 
$\approx 10^{-17} $s$=10$as$.$
Recently the attosecond laser pulses were obtained and investigated 
\cite{1}--\cite{6}. From the above discussion it is obvious that attosecond 
laser pulse create the perturbation of structure of the matter 
on the atomic level and for the fact that 
$\Delta t\leq \tau $
the equation~(\ref{eq1}) is the master equation for the thermal 
processes on the nanometer scale.

The Eq.~(\ref{eq1}) is the hyperbolic PDE. The equation with the 
same mathematical structure describes for example propagation 
of the electromagnetic field \textit{E} in the conductors~\cite{8}
    \begin{equation}
\frac{\partial ^{2} E}{\partial t^{2} } +\frac{\sigma _{0} }{\epsilon
_{0} } \frac{\partial E}{\partial t} =c^{2} \nabla ^{2} E.\label{eq3}
    \end{equation}
With the substitution 
    \begin{equation}
E=e^{-\frac{t}{2\tau } } f(x,t).\label{eq4}
    \end{equation}
Eq.~(\ref{eq3}) can be written as
    \begin{equation}
\frac{\partial ^{2} f}{\partial t^{2} } +\left[ -\frac{1}{\tau }
+\frac{\sigma _{0} }{\epsilon _{0} } \right] \frac{\partial f}{\partial t}
+f(x,t)\left[ \frac{1}{\tau ^{2} } -\frac{\sigma _{0} }{\epsilon _{0} }
\frac{1}{2\tau } \right] =c^{2} \nabla ^{2} f(x,t).\label{eq5}
    \end{equation}
The \textit{Ansatz} 
    \begin{equation}
\tau =\frac{\epsilon _{0} }{\sigma _{0} }\label{eq6}
    \end{equation}
defines the relaxation (decoherence) time for the electromagnetic 
field $E\left( \vec{r} \right) $.

Comparing Eq.~(\ref{eq2}) and (\ref{eq6}) we get the definition of the 
electrical conductivity of inneratomic space:
    \begin{equation}
\frac{\epsilon _{0} }{\sigma _{0} } =\frac{\hbar }{m\alpha ^{2} c^{2} }\label{eq7}
    \end{equation}
i.e.
    \begin{equation}
    \sigma _{0} =\epsilon _{0} \frac{m\alpha ^{2} c^{2} }{\hbar } \approx
10^{6} \frac{1}{\Omega {\rm m}}\label{eq8}
    \end{equation}
for $\epsilon _{0} \approx 8.8\cdot 10^{-12} \frac{\rm F}{\rm m} $
 in SI units, where \textit{F} means farad.

From formula~(\ref{eq8}) we conclude that conductivity of the 
space time for the atomic electron is described by the fundamental 
constants of nature. The conductivity $\sigma _{0} $, formula~(\ref{eq8}), is the pure quantum effect. Formula~(\ref{eq8}) 
can be written as
    \begin{equation}
    \sigma _{0} =\frac{\epsilon _{0} \alpha c}{\lambda _{B} } ,\label{eq9}
    \end{equation}
where $\lambda _{B} $ denotes the de Broglie'a wave length of the heaton~\cite{1}
    \begin{equation}
    \lambda _{B} =\frac{\hbar }{m\alpha c}.\label{eq10}
    \end{equation}
\section{The Green function for the Klein-Gordon thermal equation in one 
and three dimensions}
In the generalized heat transport equation~(\ref{eq1}) we seek a solution 
in the form 
    \begin{equation}
    T\left( \vec{r},t \right)=e^{-\frac{t}{2\tau } } u(\vec{r},t),\label{eq11}
    \end{equation}
where the relaxation time $\tau $ equals 
    \begin{equation}
    \tau =\frac{\hbar }{mv^{2} }.\label{eq12}
    \end{equation}
After substitution Eq.~(\ref{eq11}) into Eq.~(\ref{eq1}) one obtains  
    \begin{equation}
    \frac{\partial ^{2} u}{\partial t^{2} } -v^{2} \Delta u+q^{2} u\left(
\vec{r},t \right)=0,\label{eq13}
    \end{equation}
where 
    \begin{equation}
    q^{2} =\frac{2V_{0} mv^{2} }{\hbar ^{2} } -\left( \frac{mv^{2} }{2\hbar }
\right) ^{2}.\label{eq14}
    \end{equation}
First we solve in one dimension:
    \begin{eqnarray}
    \frac{\partial ^{2} u}{\partial t^{2} } -v^{2} u_{xx} +q^{2} u=0,\;&\quad&
-\infty <x<\infty ,\label{eq15}\\
    u(x,0)=\Phi (x), &\quad& u_{t} (x,0)=\Psi (x),\label{eq16}
    \end{eqnarray}
where \textit{v} and \textit{q} are positive constant.
 
The Fourier transform of the Green function for one dimensional 
Klein-Gordon equation fulfills the equation~\cite{9}
    \begin{equation}
    \frac{\partial ^{2} G(k,t)}{\partial t^{2} } =-v^{2} k^{2} G(k,t)-q^{2}
G(k,t)\label{eq17}
    \end{equation}
with boundary conditions 
    \begin{equation}
    G(k,0)=0,\quad \quad \frac{\partial G(k,t)}{\partial t} (k,0)=1.\label{eq18}
    \end{equation}
The Eq.~(\ref{eq4}) has the solution
    \begin{equation}
    G(k,t)=\frac{\sin \left( t\sqrt{v^{2} k^{2} +q^{2} } \right)
}{\sqrt{v^{2} k^{2} +q^{2} } }. \label{eq19}
    \end{equation}
In the (\textit{x}, \textit{t}) plane \textit{G}(\textit{x}, \textit{t}) has the form~\cite{9}\\
    \begin{eqnarray}
G(x,t)&=&\int\limits_{-\infty }^{\infty }\frac{\sin \left( t\sqrt{v^{2}
k^{2} +q^{2} } \right) }{\sqrt{v^{2} k^{2} +q^{2} } } e^{ik}
\frac{dk}{2\pi } \nonumber\\
&=&\frac{1}{2c} J_{0} \left( q\sqrt{t^{2} -\frac{x^{2}
}{v^{2} } } \right)  \quad {\rm for} \quad \left| x\right| <vt,\label{eq20}\\
 G(x,t)&=&0  \quad {\rm for} \quad \left| x\right| >vt\label{eq21}   
    \end{eqnarray}
and \textit{J}$_{0}$ is the modified Bessel function.

In three dimensions using the same method we obtain~\cite{9}
    \begin{equation}
    G\left( \vec{r}, t\right)=\iiint_{-\infty}^{\infty}\frac{\sin\left(t\sqrt{v^2k^2+q^2}\right)}
{\sqrt{v^2k^2+q^2}}e^{i\vec{k}\vec{r}}\frac{d\vec{k}}{8\pi^3}.\label{eq22}
    \end{equation}
We will calculate the integral~(\ref{eq8}) in the spherical coordinates, 
Let $\theta $ denotes the angle between $\vec{k} $
 and $\vec{r} $. Then Eq.~(\ref{eq8}) takes the form 
 \begin{equation}
 G\left( \vec{r} , t\right)=\int\limits_0^{2\pi}\int\limits_0^{\pi}\int\limits_0^{\infty}\frac{\sin\left(t\sqrt{v^2k^2+q^2}\right)}
{\sqrt{v^2k^2+q^2}}e^{ikr\cos\theta}\times\frac{k^2\cos\theta \,dk\,d\theta \, d\Phi}{8\pi^3}.\label{eq23}
    \end{equation}
The $\Phi $  and $\theta $
 integrals are easily integrated out to get
\begin{equation}
G\left( \vec{r},t \right)=-\frac{1}{4\pi^2r}\frac{\partial}{\partial r}
\int_0^{\infty}\frac{\sin \left(t\sqrt{v^2k^2+q^2}\right)}{\sqrt{v^2k^2+q^2}}
k\sin(kr)dk.\label{eq24}
\end{equation}
Now we can write 
$k\sin (kr)=\frac{\partial (-\cos (kr))}{\partial r} $
 pull the $\frac{\partial }{\partial r} $
 outside of integral and use the fact that the integral is an 
ever function of \textit{k}, to get~\cite{9}
    \begin{equation}
   G\left( \vec{r},t \right)=-\frac{1}{4\pi^2r}\frac{\partial}{\partial r}
\int_{-\infty}^{\infty}\frac{\sin \left(t\sqrt{v^2k^2+q^2}\right)}{\sqrt{v^2k^2+q^2}}
e^{ikr}dk\label{eq25}
    \end{equation}
i.e. 
    \begin{equation}
  G\left( \vec{r},t \right)=-\frac{1}{4\pi v r}\frac{\partial}{\partial r}
  \left[H\left(t^2-\frac{r^2}{c^2}\right)J_0
\left(q\sqrt{t^2-\frac{r^2}{v^2}}\right)\right].\label{eq26}
    \end{equation}
Carrying out the derivative and using the identity $J_{0} ^{} =-J_{1} $
 and $J_{0} (0)=1$
 we get~\cite{9}
    \begin{equation}
G\left( \vec{r},t \right)=\frac{1}{2\pi c}\delta\left(v^2t^2-r^2\right)
-qH\left(v^2t^2-r^2\right)\frac{J_1\left[\frac{\displaystyle q}{\displaystyle v}\sqrt{v^2t^2-r^2}\right]}
{4\pi v^2\sqrt{v^2t^2-r^2}}.\label{eq27}
    \end{equation}
This means that the Green function for thermal K-G equation in 
three dimensions is the Dirac delta function 
$\delta \left( v^{2} t^{2} -r^{2} \right) $
 on the wave cone 
$r=vt$
 plus a Bessel function inside the cone.
 
\section{Conclusions}
The borderland between the classical and quantum physics is extensively 
investigated. In this paper the quantum Klein-Gordon thermal 
(KGT) equation is discussed. The KGT is obtained when mean free 
path is equal to the de Broglie'a wave length. Due to its quantum 
nature KGT can be used to the description of the thermal processes 
on the atomic scale. In this paper the solution of KGT for one 
and three dimensional transport phenomena are obtained and discussed. 
The conductivity for space-time in the atom is calculated and 
the value 
$\sigma _{0} =10^{6} \frac{1}{\Omega {\rm m}} $
 is obtained.

\end{document}